\documentclass{iau}
\usepackage{graphicx}

\title[~~Morphological Types of 247 PF Galaxy Clusters]
{The Adapted Morphological Types \\ of 247 Rich PF Galaxy
Clusters}

\author[Elena Panko et al.]
{Elena Panko$^1$, Katarzyna Bajan$^2$, Piotr Flin$^3$ \and Alla Gotsulyak$^4$}

\affiliation{$^1$Kalinenkov Astronomical Observatory, Nikolaev National University, Nikolaev, Ukraine\\
email: {\tt panko.elena@gmail.com} \\[\affilskip]
$^2$Institute of Physics, Pedagogical University, Cracow, Poland, \\
$^3$Institute of Physics, Jan Kochanowski University, Kielce, Poland \\
$^4$ Astronomical Department, Odessa National University, Odessa, Ukraine}

\pubyear{20XX}
\volume{308}  
\pagerange{000}
 \jname{The Zeldovich Universe \\ Genesis and
Growth of the Cosmic Web}

\begin{document}

\maketitle

\begin{abstract}
Morphological types were determined for 247 rich galaxy clusters
from the PF Catalogue of Galaxy Clusters and Groups. The adapted
types are based on classical morphological schemes and consider
concentration to the cluster center, the signs of preferential
direction or plane in the cluster, and the positions of the
brightest galaxies. It is shown that both concentration and
preferential plane are significant and independent morphological
criteria.


\end{abstract}

\section{Introduction}

The classification of galaxy clusters at optical wavelengths is
carried out using several different parameters: cluster richness
(number of galaxies within a specific limiting magnitude), degree of
central concentration, the presence of bright galaxies in the
center of the cluster, etc. The prevalent Bautz-Morgan (BM)
(\cite[Bautz \& Morgan, 1970]{BM}) and Rood-Sastry (\cite[Rood \&
Sastry, 1971]{RS}) classification schemes are in agreement and
complement each other. \cite[L\'{o}pez-Cruz et al.
(1997)]{Lopez-Cruz97} introduced the definition of a cD cluster,
the complement of which is called a non-cD cluster.

\section{Observational Data}

A Catalogue of Galaxy Clusters and Groups (\cite[Panko \& Flin,
2006]{PF06}, hereafter PF) was constructed from the M{\"u}nster
Red Sky Survey Galaxy Catalogue (\cite[Ungrue, Seitter \&
Duerbeck, 2003]{Ungrue03}, hereafter MRSS) mainly for statistical
analysis of properties for large-scale structures. Unfortunately,
so far we have only been able to study the cluster parameters from
the morphology for 1056 PF clusters that are coincident with those
in the ACO catalogue (\cite[Abell, Corvin \& Olovin, 1989]{ACO}).
Similarly, only 247 PF clusters with richness $N\geq100$ have
assumed BM morphological types according to the ACO catalogue.
Those morphological types permit us to find alignments of the
brightest galaxy relative to the parent clusters for BM type I
(\cite[Panko, Juszczyk \& Flin, 2009]{PJF09}). The Binggeli effect
(\cite[Binggeli, 1982]{Bing}) is strongest for BM type I clusters,
as well presents for BM III clusters (\cite[Flin et al.,
2011]{Flin BA}). Moreover, \cite[God{\l}owski et al.
(2010)]{GodlBA}, using data for 97 PF galaxy clusters, found a
weak dependence of galaxy velocity dispersion with BM type for the
parent cluster. Other morphology schemes for PF clusters were not
used. Presently we adapted the prevalent morphological systems for
the MRSS observational data and determined our morphological types
for PF galaxy clusters using a $2D$ distribution of galaxies in
rectangular coordinates relative to the cluster center for each.

\begin{figure}[t]
\begin{center}
 \includegraphics[width=2.8in]{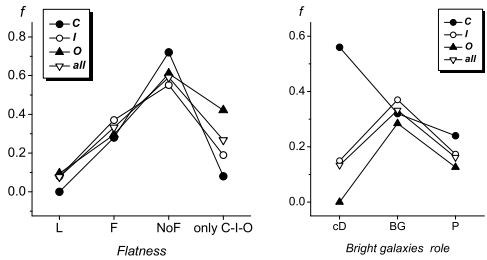}
 \caption{Variation of frequencies for flatness signs (left) and BCM role in C-I-O and all clusters. }
   \label{fig1}
\end{center}
\end{figure}

\section{The adapted morphological scheme}

For input data we established adapted morphological types based on
3 parameters: concentration, signs of flatness, and bright cluster
members (BCM) positions. The adapted types correspond to
concentration (C \-- compact, I \-- intermediate, and O \-- open),
flatness (L \-- line, F \-- flat, and no symbol if no indication
of flatness is present), and the role of bright galaxies  (cD or
BG if the BCM role is significant). Other peculiarities are noted
as P. The details of the approach are described and justified
elsewhere \cite[Panko (2013)]{P13}. The designations can be
combined, for example CFcD or ILP.

For 247 rich PF clusters with BM types from the ACO comparison, we
determined the adapted morphological types and analyzed the
frequencies of each. The sign of the flatness type is independent
of concentration class: those for L and F types are similar in
C-I-O groups, as shown in Fig.\,\ref{fig1}, left panel. In
contrast, the role of BCMs is strongly connected with cluster
concentration: the number of cD clusters is greatest in C-type
(Fig.\,\ref{fig1}, right panel).  Note, CcD type corresponds to BM
I type. For L and F clusters we found a correlation between
position angle for the major axes of the best-fit ellipse and the
direction of the preferred plane.

\section{Conclusions}

From $2D$ maps of 247 rich PF galaxy clusters we determined their
adapted morphological types. It is shown that concentration and
flatness are independent morphological criteria. The direction of
the major axis of the best-fitting ellipse for a cluster
(calculated in the PF catalogue) is close to the direction
determined by the L or F regions; the difference between the two
directions increases for O-type galaxy clusters.

\end{document}